\documentstyle[aps,preprint]{revtex}
\begin{document}
\draft
\title{An improved perturbation approach to the 2D
Edwards polymer -- corrections to scaling}
\author{S. R. Shannon, T. C. Choy and R. J. Fleming}
\address{Department of Physics, Monash University, Clayton,
Victoria, Australia}
\date{\today}
\maketitle
\begin{abstract}
We present the results of a new perturbation calculation in polymer
statistics which starts from a ground state that already correctly predicts
the long chain length behaviour of the mean square end--to--end distance
$\langle R_N^2 \rangle\ $, namely the solution to the 2~dimensional~(2D)
Edwards model. The $\langle R_N^2 \rangle$ thus calculated is shown to
be convergent in $N$, the number of steps in the chain, in contrast to
previous methods which start from the free random walk solution.
This allows us to calculate a new value for the leading correction--to--scaling
exponent~$\Delta$. Writing $\langle R_N^2 \rangle = AN^{2\nu}(1+BN^{-\Delta}
+ CN^{-1}+...)$, where $\nu = 3/4$ in 2D,  our result shows
that $\Delta = 1/2$. This value is also supported by an analysis of 2D
self--avoiding walks on the {\em continuum}.
\end{abstract}

\section{Introduction}

The central quantity of interest in polymer statistics is the mean square
end--to--end distance $\langle R^2_N\rangle$ for a chain of $N$ links.
The excluded volume effect causes a `swelling' of the chain beyond the
$\langle R^2_N\rangle \propto N$ relationship of a free random walk.
Thus
\begin{equation}
\langle R^2_N \rangle = AN^{2\nu}(1+BN^{-\Delta}+CN^{-1}+ ...)\ ,
\label{rform}
\end{equation}
in the large $N$ limit, where $\nu$ is the leading scaling exponent,
$A,B,C$ are excluded volume dependent coefficients and $\Delta$ is the
leading correction--to--scaling exponent \cite{wegner}.  It is now firmly
established  \cite{nienhuis,guttmann,ishinabe,santos} that in two
dimensions (2D) $\nu=3/4$ is exact. Despite this, there is very little
agreement about the value of $\Delta$. Nienhuis \cite{nienhuis}
predicts $\Delta=3/2$, while Rapaport \cite{rapaport} has argued
that there is no need for a correction term other than the analytic
correction, i.e., $\Delta=1$. However, many numerical studies have disagreed
with these results, with estimates for $\Delta$ of $1.2$ \cite{havlin},
 $0.84$ \cite{lyklema} and $0.65$ \cite{ishinabe,santos,privman}.
These numerical estimates are based on results obtained from self--avoiding
walks on 2D lattices. Theoretical results are also in disagreement.
Besides Nienhuis's prediction, which relies on a mapping to
an exactly solvable solid--on--solid model on the honeycomb lattice,
Baker~{\em et al\/} \cite{baker} predict \mbox{$\Delta=1.18$} using
RG arguments, while Saleur \cite{saleur} predicts
\mbox{$\Delta=11/16$} by conformal invariance. Interestingly, Saleur also
gives evidence for $\Delta$ but he then rejects this result.
Perturbation expansion techniques \cite{fixman,muthukumar}, which start from
the free random--walk solution, have also been used to predict
$\langle R^2_N\rangle$, but these methods have resulted in series
which are divergent in $N$ and $v$, the excluded volume parameter, and hence
a value for $\Delta$ cannot be predicted.
The obvious confusion in both the numerical and theoretical estimates
for $\Delta$ motivates the search for a better perturbation procedure as
discussed here.

In this paper we report results based on a new perturbation method,
which unlike previous studies, starts from a ground state that already
correctly predicts the {\em exact\/} large $N$ behaviour in 2D, namely
the Edwards self--consistent solution \cite{edwards}. Our approach mimics
the cluster integral method in statistical mechanics \cite{huang}.
Although it has been shown \cite{fisher}
that the Edwards solution cannot be the correct form for the
self--avoiding random walk end--to--end distribution function \cite{bishop},
it does predict an {\em exact\/} exponent $\nu$ in 2D, although
not in 3D. As far as we know, the reason for the coincidence of $\nu$ in 2D
predicted by the self--consistent theory with the recently proved
exact value \cite{saleur} is not well understood.
Thus it appears that there is essential physics in the 2D Edwards
solution that is not present in 3D, as easily seen via a simple
dimensional argument (see section IV). Mathematically the Edwards solution
also has convenient features that we can exploit in a novel
perturbation expansion.

In section II, we spell out the method for the approach. Here the essential
physics consists of (a) perturbation about a new ground state, i.e., about the
Edwards solution which is known to predict an exact exponent $\nu=3/4$
in 2D, and (b) a new perturbation potential $V(r)$, the latter being much
better
behaved than the bare excluded volume potential (see Eq.\ (\ref{our_ground})).
In section III, we present details of the calculation and results and in
section
IV we conclude with some discussions.

\section{Method}

We start from the path integral representation of the exact end--to--end
distance distribution function for the Edwards model \cite{wiegel}.
This is given by
\begin{equation}
G({\rm{\bf R}},L) = \int^{{\rm{\bf r}}(L)={\rm{\bf R}}}_{{\rm{\bf r}}(0)
={\rm{\bf 0}}}
D[{\rm{\bf r}}]\exp\left(-\frac{1}{l}\int^L_0ds
(\frac{\partial {\rm{\bf r}}(s)}{\partial s})^2 -
\omega\int^L_0ds\int^L_sds'\delta^2[{\rm{\bf r}}(s)-
{\rm{\bf r}}(s')]\right) \ ,
\end{equation}
where $L$ is the total chain length $L=Nl$, $l$ being the step length of one
link, and $\omega=v/l^2$, where $v$ is the excluded `volume'. It is known that
\begin{equation}
v=\int[1-e^{-u({\rm{\bf r}}_{ij})/k_BT}]d^2r_{ij} \ ,
\end{equation}
where $u({\rm{\bf r}}_{ij})$ is the pair potential between
the $i$the and $j$th segment,
$k_B$ is the Boltzmann constant and $T$ the temperature.  Traditionally
two problems arise in dealing with this intractable path integral. Firstly,
divergences appear in the calculation which must be handled carefully, and
secondly, the resulting series expansion is a power series of increasing $L$
and $\omega$. This leads to a result whose radius
of convergence for $\omega$ diminishes to zero as $L$ increases.
This property is the hallmark of modern
critical phenomena theory, whose resolution was offered by the renormalization
group approach \cite{skma,amit}. Historically Edwards avoided
the divergence problems of such an approach by replacing
the point contact potential by a self--consistent field $W(r)$ which
in 2D is equal to $v\tilde{p}(r)/l^2$,\, where $\tilde{p}(r)$ is the
one--particle potential proportional to $r^{-2/3}$ \cite{edwards}.
Therefore
\begin{equation}
W(r) = {\cal C}v^{2/3}r^{-2/3}\ ,
\end{equation}
where ${\cal C}=(\sqrt{3}/4\pi)^{2/3}l^{-5/3}$. Thus the Edwards
Green's function $G_E({\rm{\bf R}},L)$ becomes
\begin{equation}
G_E({\rm{\bf R}},L)=\int^{{\rm{\bf r}}(L)={\rm{\bf R}}}_{{\rm{\bf r}}(0)
={\rm{\bf 0}}}D[{\rm{\bf r}}]
\exp\left(-\frac{1}{l}\int^L_0ds\left(\frac{\partial {\rm{\bf r}}(s)}{\partial
s}\right)^2 -
\int^L_0 W(s) ds \right)\ .
\label{ed_green}
\end{equation}
Our approach relies on obtaining a better
first order perturbation expansion by starting from the Edwards ground state
and then perturbing this with the {\em difference\/} between the
self--consistent field and the true point contact potential defined
by $V(s)$ below. Thus
\begin{eqnarray}
G({\rm{\bf R}},L) &=& G_E({\rm {\bf R}},L) \times \exp\left( \int^L_0 W(s) ds -
\omega\int^L_0ds\int^L_sds'\delta^2[{\rm{\bf r}}(s)-
{\rm{\bf r}}(s')]\right) \nonumber \\
&\equiv& G_E({\rm{\bf R}},L) \times \exp\left(\int_0^L V(s)\,ds\right)\ .
\label{our_ground}
\end{eqnarray}
To a first order perturbation expansion in $V(s)$, this now becomes
\begin{eqnarray}
G({\rm{\bf R}},L) &=& G_E({\rm{\bf R}},L)\, \times
\left[ 1 + \int_0^L V(s)\,ds + {\cal O}(V(s)^2) \right]\ , \nonumber \\
&=& G_E({\rm{\bf R}},L) + G_V({\rm{\bf R}},L) + ...\ .
\label{solution}
\end{eqnarray}
The Fourier transform of Eq.\ (\ref{solution}) is
\begin{eqnarray}
\hat{G}({\rm{\bf k}},L) &=& \hat{G}_E({\rm{\bf k}},L) +
\hat{G}_V({\rm{\bf R}},L) + ...\ , \nonumber \\
&=& \hat{G}_E({\rm{\bf k}},L) + \hat{G}_1({\rm{\bf k}},L)
+ \hat{G}_2({\rm{\bf k}},L) + ...\  ,
\label{series}
\end{eqnarray}
where $\hat{G}_E({\rm{\bf k}},L)$ is the Fourier transform of the Edwards
Green function, $\hat{G}_1({\rm{\bf k}},L)$ the transform of the
self--consistent field acting on  $\hat{G}_E({\rm{\bf k}},L)$, and
$\hat{G}_2({\rm{\bf k}},L)$ is the transform of the point contact potential
acting on $\hat{G}_E({\rm{\bf k}},L)$.
If we now define the following Laplace transformation function
\begin{equation}
\tilde{G}_E({\rm{\bf k}},E)  =  \int^\infty_0 dL\:
e^{-EL}\hat{G}_E({\rm{\bf k}},L) =
{\cal L}\{\hat{G}_E({\rm{\bf k}},L)\}\ ,
\end{equation}
then it can be shown \cite{muthukumar} that due to the properties of the
delta function in Eq.\ (\ref{solution}) and the factorizability of
the path integral of Eq.\ (\ref{ed_green})
\begin{equation}
\tilde{G}_2({\rm{\bf k}},E) = \tilde{G}_E({\rm{\bf k}},E)^2
\int\frac{d^2\!p}{4\pi^2}\tilde{G}_E(p,E)\ .
\label{old_g2}
\end{equation}
At this stage we note a simplification over the method in ref.
\cite{muthukumar}
which bypasses the need to evaluate the integral of $\tilde{G}_E(p,E)$ as it is
non--trivial in our case. This can be seen by noting that
\begin{eqnarray}
\int\! \frac{1}{4\pi^2} \tilde{G}_E({\rm{\bf p}},E)\,d^2\!p &=&
\int\!\frac{1}{4\pi^2}\int e^{i{\rm{\bf p}}.{\rm{\bf R}}}\,\hat{G}_E
({\rm{\bf R}},E)\:d^2\!R\,d^2\!p\ , \nonumber \\
 &=& \int\!\! \int\!\frac{1}{4\pi^2} e^{i{\rm{\bf p}}.{\rm{\bf R}}}\,d^2\!p\,
\hat{G}_E({\rm{\bf R}},E)\:d^2\!R\ , \nonumber \\
 &=& \int \delta^2({\rm{\bf R}})\,\hat{G}_E({\rm{\bf R}},E)\:d^2\!R\ ,
\nonumber \\
 &\equiv& \bar{G}_E(0,E)\ .
\end{eqnarray}
Thus $\tilde{G}_2({\rm{\bf k}},E)$ is given by
\begin{eqnarray}
\tilde{G}_2({\rm{\bf k}},E) &=& \tilde{G}_E({\rm{\bf k}},E)^2\bar{G}_E(0,E)\ ,
\nonumber \\
 &=& \tilde{G}_E({\rm{\bf k}},E)^2{\cal L}
\{G_E({\rm{\bf R}},L)_{{\rm{\bf R}}=0}\}\ .
\label{g2_calc}
\end{eqnarray}
Using the convolution property of the Laplace transform, then
\begin{eqnarray}
\hat{G}_2({\rm{\bf k}},L) &=& {\cal L}^{-1}\!\left[\tilde{G}_E({\rm{\bf
k}},E)^2
{\cal L} \{G_E({\rm{\bf R}},L)_{{\rm{\bf R}}=0}\}\right]\ , \nonumber  \\
   &=& \int^L_0 {\cal L}^{-1}\!\left[\tilde{G}_E({\rm{\bf k}},L-u)^2\right]
(G_E({\rm{\bf R}},u)_{{\rm{\bf R}}=0})\:du \ ,
\label{convolution}
\end{eqnarray}
which is a convenient result. Once $\hat{G}_2({\rm{\bf k}},L)$ is obtained
from (\ref{convolution}) we can then calculate the mean
square end--to--end chain distance by
\begin{eqnarray}
\langle R^2 \rangle & = & \int d^2\!R\, R^2G({\rm{\bf R}},L)\left/
\int d^2\!R\, G({\rm{\bf R}},L)\right.\ , \nonumber \\
 & = & -4\left[\frac{\partial}{\partial k^2}
\hat{G}({\rm{\bf k}},L)\right]_{{\rm{\bf k}}=0}\left/\hat{G}(0,L)\right.\ .
\label{mean_calc}
\end{eqnarray}
The formula (\ref{mean_calc}) overcomes the need to calculate the integral
in Eq.\ (\ref{old_g2}) which is non--trivial given the form
of $\tilde{G}_E(p,E)$. Moreover, it is worth noting that the above
calculations are only tractable in our case because of the
essentially (shifted) Gaussian nature of the 2D Edwards solution
and the trick given in Eq.\ (\ref{convolution}). It appears that
a similar method is not possible with more sophisticated ground states
\cite{bishop}.

\section{Calculations and Results}

We shall first show that the above method works by following the
traditional approach and starting from the free walk solution in 2D. Given that
\begin{equation}
G_0({\rm{\bf R}},L) = \frac{1}{\pi lL}\exp(-\frac{1}{lL}{R^2})\ ,
\end{equation}
thus
\begin{equation}
{\cal L}\{G_0({\rm{\bf R}},L)_{{\rm{\bf R}}=0}\} =
{\cal L}\left\{\frac{1}{\pi lL}\right\}\ ,
\label{old_ground}
\end{equation}
and
\begin{equation}
\tilde{G}_0({\rm{\bf k}},E) = \frac{1}{E+\frac{l}{4}k^2}\ .
\label{old_gr_trans}
\end{equation}
Using Eq.\ (\ref{g2_calc}), Eq.\ (\ref{old_ground})
and Eq.\ (\ref{old_gr_trans}) in Eq.\ (\ref{mean_calc}) we obtain
\begin{equation}
\langle R^2 \rangle = \frac{{\cal L}^{-1}\left[\frac{l}{E^2}\right]
- {\cal L}^{-1}\left[\frac{2l\omega}{E^3}{\cal L}
\left\{\frac{1}{\pi lL}\right\}\right]} {{\cal L}^{-1}\left[\frac{1}{E}\right]
- {\cal L}^{-1}\left[\frac{\omega}{E^2}{\cal L}\left\{\frac{1}{\pi lL}\right\}
\right] }\ .
\end{equation}
Introducing a cutoff $\epsilon$ at the small $L$ limit, this reduces to
\begin{eqnarray}
\langle R^2 \rangle &=& \frac{lL - \frac{\omega}{\pi}[L^2\log(L) -
\frac{3}{2}L^2 - L^2\log(\epsilon)]}{1 - \frac{\omega}{\pi l}[L\log(L) - L
-L\log(\epsilon)]}\ , \nonumber \\
&=& lL\left(1+ \frac{\omega}{2\pi l}L + {\cal O}(\omega^2)\right)\ .
\end{eqnarray}
which agrees with the previously known result \cite{fixman,muthukumar}.

It can be shown \cite{edwards} that in 2D the Edwards
solution for the probability distribution for the end--to--end
distance for a chain in the large $L$ limit,
i.e., the solution of Eq.\ (\ref{ed_green}), is given by
\begin{equation}
G_E({\rm{\bf R}},L) = N(L) \exp(-\frac{B}{L}({\rm{\bf R}}-AL^{3/4})^2)\ ,
\label{ground}
\end{equation}
where
\begin{equation}
A=\left(\frac{2^7}{3^4\pi}\right)^{1/4}\!\left(\frac{v}{l}\right)^{1/4}\!,
 \;\;\;  B=\frac{9}{8l}\ .
\label{ab_values}
\end{equation}
and $N(L)$ is the normalisation. Unlike Edwards \cite{edwards}, the
normalisation is needed in our calculation. This is determined by
the identity
\begin{equation}
\int G_E({\rm{\bf R}},L) d^2{\rm{\bf R}} = 1\ .
\label{norm_eq}
\end{equation}

Given the asymptotic form of $G_E({\rm{\bf R}},L)$ in (\ref{ground}) we can
calculate $N(L)$ from (\ref{norm_eq}), see Appendix A. The result is
\begin{equation}
N(L) = 1\left/\left( \frac{2A\pi^{3/2}}{\sqrt{B}}L^{5/4} +
	\frac{\pi}{2A^2B^2}L^{1/2}e^{-A^2B\sqrt{L}}
\right)\right. \ .
\label{normalisation}
\end{equation}
For later reference we note that the exponentially decreasing term in the
denominator of the above expression for $N(L)$ appears as a natural `cutoff'
and prevents many subsequent integrals
from diverging. It plays a similar role as the cutoff $\epsilon$,
but does not need to be artificially introduced as in the perturbation
expansion based on the free--walk solution given above.

{}From Eq. (\ref{mean_calc}) it can be seen that only the small {\bf k}
behaviour is required. Thus the Fourier transform of $G_E({\rm{\bf R}},L)$,
see Appendix B, becomes
\begin{equation}
\hat{G}_E({\rm{\bf k}},L) = 1 - \frac{k^2}{4}\left[ A^2 L^{3/2}
+ \frac{3}{2B}L \right]\ .
\label{g_e}
\end{equation}
The Laplace transform of the above gives
\begin{eqnarray}
\tilde G_E({\rm{\bf k}},E) &=& {\cal L} \{ 1 -\frac{k^2}{4}[A^2L^{3/2}
+\frac{3}{2B}L\,]\}\ , \nonumber \\
&=& \frac{1}{E} - \frac{k^2}{4}(A^2 \frac{3\sqrt{\pi}}{4}E^{-5/2} +
\frac{3}{2B}
E^{-2})\ .
\label{laplace}
\end{eqnarray}

By examining Eq. (\ref{solution}) and the form of the one--particle potential
$W(r)$, it is seen that the
term $G_1({\rm{\bf R}},L)$ is calculated simply by
differentiation of $G_E({\rm{\bf R}},L)$ w.r.t. the excluded volume parameter.
Thus
\begin{eqnarray}
\hat{G}_1({\rm{\bf k}},L) &=&
-v^{2/3}\frac{\partial}{\partial(v^{2/3})}\hat{G}_E({\rm{\bf k}},L)\ ,
\nonumber \\
&=& -v^{2/3}\frac{\partial}{\partial(v^{2/3})}\left( 1 -
\frac{k^2}{4}[A^2L^{3/2} + \frac{3}{2B}L\,]\right)\ .
\label{g1_calc}
\end{eqnarray}
The higher order terms present all decay exponentially and thus are
insignificant in the large $L$ limit. Since only $A$ has a $v$ dependence,
then from Eq. (\ref{ab_values})
\begin{eqnarray}
\hat G_1({\rm{\bf k}}, L) = \frac{3}{4} \frac{\,k^2}{4}A^2L^{3/2}\ .
\label{g_1}
\end{eqnarray}
We note this is just $-\frac{3}{4}G_E({\rm{\bf k}},L)$ to leading order in $L$,
and it acts to decrease the value of~$\langle R^2_N \rangle$.
The term $\hat{G}_2({\rm{\bf k}},L)$ is calculated using Eq.\ (\ref{g2_calc}),
Eq.\ (\ref{ground}) and Eq.\ (\ref{laplace}) to order $k^2$, thus
\begin{equation}
\tilde{G}_2({\rm{\bf k}},E) = \left[ \frac{1}{E^2} - \frac{k^2}{4}\left(
A^2\frac{3\sqrt{\pi}}{2}E^{-7/2} + \frac{3}{B}E^{-3}\right)\right]{\cal L}
\{N(L)\exp(-A^2B\sqrt{L}\,)\}\ .
\end{equation}
Now
\begin{equation}
{\cal L}^{-1}\left\{ \frac{1}{E^2} -
\frac{k^2}{4}\left(A^2\frac{3\sqrt{\pi}}{2}E^{-7/2} + \frac{3}{B}
E^{-3}\right) \right\} =
L - \frac{k^2}{4}\left( \frac{4}{5}A^2L^{5/2} + \frac{3}{2B}L^2\right)\ ,
\end{equation}
and
\begin{equation}
{\cal L}^{-1}\left\{ {\cal L}\left[ N(L)\exp(-A^2B\sqrt{L}\,)\right]\right\} =
N(L)\exp(-A^2B\sqrt{L}\,)\ .
\end{equation}
Using the convolution property, Eq. (\ref{convolution}), we arrive at
\begin{equation}
\hat{G}_2({\rm{\bf k}},L) = \int_0^L \left[ (L-u) - \frac{k^2}{4}\left(
\frac{4}{5}A^2(L-u)^{5/2} + \frac{3}{2B}(L-u)^2\right) \right]
N(u)\exp(-A^2B\sqrt{u}\,)\, du\ .
\label{g2_equation}
\end{equation}
By examining Eq. (\ref{g2_equation}) we see that there are 3 integrals of
the form
\begin{equation}
I_s = c\int^L_0 (L-u)^s N(u) \exp(-A^2B\sqrt{u})\,du\ ,
\end{equation}
with $s=1$, $s=5/2$ and $s=3$. Writing $N(u)$ from Eq.\ (\ref{normalisation})
as follows
\begin{equation}
N(u) =
1\left/ \left(M u^{5/4} + T u^{1/2}\exp(-A^2B\sqrt{u}\,)\right)\right.\ ,
\end{equation}
where $M=2A\pi^{3/2}/\sqrt{B}$ and $T=\pi/2A^2B^2$. Simple manipulations ( see
Appendix C ) lead to the result
\begin{equation}
I_s = 2cL^s\Phi_0 - 2csL^{s-1}\Phi_2 + {\cal O}(L^{s-2}\Phi_4) + ...\ ,
\end{equation}
where the integral $\Phi_p$, which has some finite value in the large $L$
limit, is given by
\begin{equation}
\Phi_p = \int_0^{\sqrt{L}}\frac{x^p}{T+Mx^{3/2}\exp(A^2Bx)}\,dx\ .
\end{equation}
Using these formulas we derive
\begin{eqnarray}
I_1 &=& 2L\Phi_0 - 2\Phi_2\ , \nonumber \\
I_{5/2} &=& -k^2\left(\frac{2A^2\Phi_0}{5}L^{5/2}
- A^2\Phi_2L^{3/2} + ... \right)\ , \nonumber \\
I_2 &=& -k^2\left(\frac{3\Phi_0}{4B}L^2 - \frac{2\Phi_2}{2B}L + ... \right)\ .
\label{integrals}
\end{eqnarray}
Using Eqs.\ (\ref{integrals}) in Eq.\ (\ref{g2_equation}), and keeping
only the leading order terms in $L$, leads to the result
\begin{equation}
\hat{G}_2({\rm{\bf k}},L) = 2L\Phi_0 - 2\Phi_2
-k^2\left[ \frac{2A^2\Phi_0}{5}L^{5/2} + \frac{3\Phi_0}{4B}L^2
 - A^2\Phi_2 L^{3/2} - \frac{3\Phi_2}{2B}L\, \right]\ .
\label{g_2}
\end{equation}
By substituting (\ref{g_e}),\ (\ref{g_1}) and (\ref{g_2}) in Eq.
(\ref{mean_calc}) we find
\begin{eqnarray}
\langle R^2 \rangle &=& -4 \frac{ -\frac{1}{4}\left(A^2L^{3/2}+
\frac{3}{2B}L\right) + \frac{3}{16}A^2L^{3/2} -\omega
\left(-\frac{2A^2\Phi_0}{5}
L^{5/2} - \frac{3\Phi_0}{4B}L^2 + A^2\Phi_2 L^{3/2} + \frac{3\Phi_2}{2B}L
\right) }
{ 1 - \omega(2L\Phi_0 - 2\Phi_2) }\ , \nonumber \\
&=& \frac{ -\frac{8}{5}A^2\Phi_0\omega L^{5/2} - \frac{3\omega\Phi_0}{B}L^2
+ \left(4A^2\omega\Phi_2 + \frac{A^2}{4}\right)L^{3/2} +
\left( \frac{6\omega\Phi_2}{B} + \frac{3}{2B}\right)L }
{ -2\omega\Phi_0 L + 2\omega\Phi_2 + 1}\ .
\end{eqnarray}
Here the integrals are all bounded and hence we can take the large
$L$ limit of the entire expression freely. Cancelling out the corresponding
$-2\omega\Phi L$ term we get
\begin{eqnarray}
\langle R^2 \rangle &=& \frac{4}{5}A^2L^{3/2} + \frac{3}{2B}L -
\left( \frac{2A^2\Phi_2}{\Phi_0} +
\frac{A^2}{8\omega\Phi_0}\right) L^{1/2} + ...\ ,
\nonumber \\
&=& \frac{4}{5}A^2L^{3/2}\left( 1 + \frac{15}{8A^2B}L^{-1/2} -
\frac{20\omega\Phi_2+\frac{5}{4}}{8\omega\Phi_0} L^{-1} + ... \right)\ .
\label{final}
\end{eqnarray}
which is our final result. We see that in deriving this result,
as no divergences are encountered,
there is  no restriction on the value of
$w$ for the convergence of the series in the large $L$ limit.
Comparing Eq.\ (\ref{final}) with Eq.\ (\ref{rform}), we see that
$\Delta=1/2$ and the coefficient $C$ is negative.
However, unless the calculation is taken to the next order,
we cannot be certain about the sign of $C$ as there are additional
contributions for this term to ${\cal O}(V(s)^2)$. We note, however, see
section IV, that a negative $C$ is consistent with our Monte Carlo simulations.

\section{Discussion}

The prediction $\Delta =1/2$, the main result of this paper, differs from all
previous predictions with the exception of that of Saleur \cite{saleur} who
rejected it for technical reasons.
We summarise the various predictions for $\Delta$
in 2D in Table (\ref{delta_vals}). The numerical simulations are of
particular interest. With the exception of a very few authors
\cite{croxton,kremer} (who however were not concerned with the
correction--to--scaling terms), it appears few simulation studies have been
done on the continuum, most previous work concentrating on 2D and 3D lattices.
Simulations of chains in the continuum is more demanding computationally
than for a lattice. We have addressed this problem with the use of a
{\em biased} sampling Monte Carlo method \cite{rosenbluth,binderbook}
to create 2D self--avoiding walks on the continuum, and have applied a finite
size scaling analysis \cite{privandfish} to the resulting data.
We have found $\Delta=1/2$ and a negative $C$ coefficient, consistent with
Eqn. (\ref{final}). We note some inconsistencies with lattice simulations
in view of the universality of $\Delta$ \cite{barma}, indicating that
the analysis of this data is strongly affected by assumptions
regarding $C$ in Eq.\ (\ref{rform}). These details will be
published elsewhere \cite{our_prl}.

Although the above calculation can also be performed starting from the
solution to the 3D Edwards model \cite{edwards}, resulting in a
convergent series in $N$, it is unlikely the correction term thus calculated
would be reliable. We provide the following heuristic argument
to support this. Since the 3D Edwards model prediction
$\nu = 3/5$ is now known to be inexact \cite{madras},  one can
easily show from a simple dimensional analysis that for
$[r] \sim L^{\nu+\delta}$, where   $\nu=3/(d+2)$, our
perturbation potential $\int_0^L V(s)\,ds$ in Eq.\ (\ref{our_ground})
varies as $L^{(4-d)/(d+2)}(\alpha_1L^{-\delta(2d-2)/3}-\alpha_2L^{-d\delta})$.
Since $\nu=3/4$ is {\em exact\/} in 2D, i.e., $\delta=0$, it is possible,
provided the amplitudes $\alpha_1=\alpha_2$
\footnote{This statement requires the equality of the {\em exact\/} amplitudes,
$\alpha_1$ and $\alpha_2$, as clearly the `perturbative' (i.e. approximate)
amplitudes cancel by virtue of the convergence of Eq.\ (\ref{final}) and
its 3D equivalent. We are unfortunately not able to prove this at present.},
for there to be a cancellation of divergences to all orders, resulting in
controlled errors in the perturbation series. This, however, cannot be
true in 3D, since $\delta<0$, even if the amplitudes are equal.

In conclusion, we have presented the results of a new perturbation method
which starts from the Edwards 2D self--consistent solution. As the later
already correctly predicts the exact exponent $\nu=3/4$ for
$\langle R^2_N \rangle$, the resulting perturbation series is shown to
be convergent in $N$ and free from uncontrolled terms.
We find the leading correction--to--scaling exponent
$\Delta=1/2$, and the next order coefficient $C$ to be negative.
These values are supported by an analysis of self--avoiding walks
on the 2D continuum \cite{our_prl}.
It appears that $\Delta=1/2$ is also
possible on the 2D lattice after a reassessment \cite{our_prl}
of current lattice data analysis methods, and that there is new physics
in the 2D Edwards solution.

TCC would like to thank Prof. Sir S. F. Edwards, Prof. D. Sherrington,
Prof. R. Stinchcombe and Prof. M. Barma for helpful discussions. SRS
acknowledges the support of an Australian Postgraduate Research Award.

\appendix
\section{Calculation of $N(L)$}
{}From Eq.\ (\ref{ground}) we have
\begin{eqnarray}
\int G_E({\rm{\bf R}},L) d^2{\rm{\bf R}}
&=& \int_0^{2\pi}\int_0^{\infty}
N(L)\exp\left(-\frac{B}{L}(R-AL^{3/4})^2\right)
R\ dR\ d\theta\ , \nonumber \\
&=& 2\pi N(L)\int_0^\infty R \exp\left(-\frac{B}{L}(R-AL^{3/4})^2\right) dR\ .
\label{our_ge}
\end{eqnarray}
Let $x=R-AL^{3/4}$
\begin{eqnarray}
&=& 2\pi N(L) \int_{-AL^{3/4}}^\infty (x+AL^{3/4})e^{-\frac{B}{L} x^2} dx\ ,
\nonumber \\
&=& 2\pi N(L) \left[
\frac{L}{2B}e^{-A^2B\sqrt{L}} + \frac{A\sqrt{\pi}}{2\sqrt{B}}
L^{5/4}\left(1+{\rm Erf}(A\sqrt{B}L^{1/4})\right)\right]\ .
\end{eqnarray}
For large $x$, the Erf($x$) is given by
\begin{eqnarray}
{\rm Erf}(x) = 1 - \frac{e^{-x^2}}{\sqrt{\pi}x} +
\frac{e^{-x^2}}{2\sqrt{\pi}x^3} -...\ .
\end{eqnarray}
Keeping the leading order terms, (\ref{our_ge}) becomes
\begin{equation}
2\pi N(L) \left[ \frac{L}{2B}e^{-A^2B\sqrt{L}} +
\frac{A\sqrt{\pi}}{\sqrt{B}}
L^{5/4} - \frac{L}{2B}e^{-A^2B\sqrt{L}} + \frac{L^{1/2}}{2A^2B^2}
e^{-A^2B\sqrt{L}} \right]\ .
\end{equation}
As noted in Eq.\ (\ref{norm_eq}), this simply equals $1$. Thus
\begin{equation}
N(L) = 1\left/\left( \frac{2A\pi^{3/2}}{\sqrt{B}}L^{5/4} +
	\frac{\pi}{2A^2B^2}L^{1/2}e^{-A^2B\sqrt{L}} \right)\right. \ .
\end{equation}

\section{ Calculation of $\hat{G}_E({\rm{\bf k}},L)$ }
{}From Eq.\ (\ref{ground}) we have
\begin{eqnarray}
\hat{G}_E({\rm{\bf k}},L) &=& \int e^{i{\rm{\bf k}}.{\rm{\bf R}}}N(L)
\exp(-\frac{B}{L}\left({\rm{\bf R}}-AL^{3/4})^2\right) d^2{\rm{\bf R}}\ ,
\nonumber \\
&=&\int_0^{2\pi}\int_0^\infty e^{ikR\cos \theta}N(L)\exp \left( -\frac{B}{L}
(R-AL^{3/4})^2 \right) R dR d\theta\ , \nonumber \\
&=& 2\pi N(L) \int_0^\infty R J_0(kR) \exp\left(-\frac{B}{L}(R-AL^{3/4})^2
\right) dR\ .
\end{eqnarray}
To order $k^2$, $J_0(kR) = 1-\frac{k^2R^2}{4}$, thus
\begin{eqnarray}
\hat{G}_E({\rm{\bf k}},L) &=& 2\pi N(L)\int_0^\infty
R(1-\frac{k^2R^2}{4})\exp\left( -\frac{B}{L}( R-AL^{3/4})^2 \right)
dR\ , \nonumber \\
&=& 2\pi N(L) \int_0^\infty R\exp\left(-\frac{B}{L}(R-AL^{3/4})^2\right) dR
\nonumber \\
&& - k^2 \frac{\pi}{2}N(L)\int_0^\infty R^3\exp\left(-\frac{B}{L}(R-AL^{3/4})^2
\right) dR\ .
\end{eqnarray}
The first part of integral simply equals 1, see Eq. (\ref{norm_eq}), while the
second $k^2$ dependent part equals
\begin{equation}
-k^2\frac{\pi}{2}N(L)\int_0^\infty R^3\exp\left(-\frac{B}{L}(R-AL^{3/4})^2
\right) dR\ .
\end{equation}
Let $x=R-AL^{3/4}$ and the above integral becomes
\begin{eqnarray}
&& -k^2\frac{\pi}{2}N(L)\int_{-AL^{3/4}}^\infty (x^3 + 3AL^{3/4} + 3A^2L^{3/2}
+ A^3L^{9/4}) e^{-\frac{B}{L}} dx\ , \nonumber \\
&=& -k^2 \frac{\pi}{2}N(L) \left[
\frac{L^2(1+A^2B\sqrt{L})}{2B^2e^{A^2B\sqrt{L}}} +
\frac{3AL^{9/4}\sqrt{\pi}}{4B^{3/2}}\left(1+{\rm Erf}(A\sqrt{B}L^{1/4})
\right)\right. \nonumber\\
&& \left. - \frac{3A^2L^{5/2}}{2Be^{A^2B\sqrt{L}}} +
\frac{3A^2L^{5/2}}{2Be^{A^2B\sqrt{L}}} +
\frac{A^3L^{11/4}\sqrt{\pi}}{2\sqrt{B}}\left(1+{\rm Erf}(A\sqrt{B}L^{1/4})
\right) \right]\ .
\end{eqnarray}
On expanding the error function, several leading terms cancel leaving
\begin{equation}
k^2\frac{\pi}{2}N(L)\left[ \frac{A^3\pi L^{11/4}}{\sqrt{B}} +
\frac{3A\sqrt{\pi}L^{9/4}}{2B^{3/2}} +
\frac{3L}{8A^4B^4}e^{-A^2B\sqrt{L}} - ... \,\right]\ . \nonumber
\end{equation}
Thus
\begin{equation}
\hat G_E({\rm{\bf k}},L) = 1 -
k^2\frac{\pi}{2}N(L)\left[ \frac{A^3\pi L^{11/4}}{\sqrt{B}} +
\frac{3A\sqrt{\pi}L^{9/4}}{2B^{3/2}} +
\frac{3L}{8A^4B^4}e^{-A^2B\sqrt{L}} - ... \,\right]\ .
\end{equation}
Substituting $N(L)$ into the above
and dropping exponentially small terms in the large $L$ limit, we get
\begin{eqnarray}
\hat{G}_E({\rm{\bf k}},L) &=& 1 - \frac{\pi}{2}\frac{\sqrt{B}}{2A\pi^{3/2}}
L^{-5/4}\left[ \frac{A^3\sqrt{\pi}}{\sqrt{B}}L^{11/4} +
\frac{3A\sqrt{\pi}}{2B^{3/2}}L^{9/4}\right] k^2\ , \nonumber \\
&=& 1 - \frac{k^2}{4}\left[ A^2 L^{3/2} + \frac{3}{2B}L \right]\ .
\end{eqnarray}

\section{ Calculation of the integral $I_s$ }
The general form for the required integral is given by
\begin{equation}
I_s = c\int_0^L\frac{(L-u)^s\exp(-A^2B\sqrt{u}\,)}
{Mu^{5/4}+Tu^{1/2}\exp(-A^2B\sqrt{u}\,)}\,du\ .
\label{gform}
\end{equation}
By a series of simple substitutions we can simplify (\ref{gform}) to
\begin{eqnarray}
I_s &=& 2cL^s \int_0^{\sqrt{L}} \frac{(1-x^2/L)^s}{T+Mx^{3/2}\exp(A^2Bx)}\, dx
\ , \nonumber \\
&=& 2cL^s\left[\,\int_0^{\sqrt{L}}\frac{1}{T+Mx^{3/2}\exp(A^2Bx)}\,dx -
\frac{s}{L} \int_0^{\sqrt{L}}\frac{x^2}{T+Mx^{3/2}\exp(A^2Bx)}\,dx\,
+ ...\, \right]\ , \nonumber \\
&=& 2cL^s\Phi_0 -2csL^{s-1}\Phi_2 + {\cal O}(L^{s-2}\Phi_4) + ...\ ,
\end{eqnarray}
where the integral $\Phi_p$ is given by
\begin{equation}
\Phi_p = \int_0^{\sqrt{L}} \frac{x^p}{T+Mx^{3/2}\exp(A^2Bx)}\,dx\ ,
\end{equation}
and is bounded for all $L$.

\begin{table}
\caption{ Predicted values for $\Delta$ in 2D \label{delta_vals}}
\begin{tabular}{rcl}
Reference & Method\tablenote{MC = Monte Carlo, FSS = Finite size scaling}
& Predicted $\Delta$ \\ \hline
Nienhuis \cite{nienhuis} & Hexagonal lattice mapping & 3/2  \\
Chaves and Riera \cite{chaves} & Cell renormalization group & 1  \\
Saleur   \cite{saleur} & Conformal invariance
& 11/16, 1/2\tablenote{This result was rejected.} \\
Havlin and Ben-Avraham \cite{havlin} & MC\tablenote{Square lattice
data using extrapolation of series analysis}
& 1.2 $\pm$ 0.1  \\
Rapaport \cite{rapaport} & MC\tablenote{Square and triangular
lattice data using linear regression} & 1  \\
Wang \cite{wang} & Exact triangular\tablenote{Fitting to a series analysis} &
0.85 $\pm$ 0.05 \\
Lyklema and Kremer \cite{lyklema} & MC\tablenote{Square lattice data
using extrapolation and fitting to a series analysis}  & 0.84  \\
Djordjevic {\em et al} \cite{santos}&
Exact triangular\tablenote{Extrapolation and fitting
to a series analysis}  & 0.66 $\pm$ 0.07 \\
Privman \cite{privman} & Exact triangular\tablenote{Intersection of a FSS
series analysis} & 0.65 $\pm$ 0.08  \\
Ishinabe \cite{ishinabe} & Exact square\tablenote{Intersection of a FSS
series analysis} & 0.65 $\pm$ 0.05 \\
Lam \cite{lam} & MC\tablenote{Square and triangular lattice using integration
of
a FSS series} & 0.6 \\
This work & Improved perturbation & 1/2 \\
Shannon {\em et al} \cite{our_prl} &
MC continuum\tablenote{Fitting of a FSS series analysis} & 0.5 $\pm$ 0.05 \\
\end{tabular}
\end{table}

\end{document}